\documentclass{PoS}
\usepackage{amsmath}

\title{The role of theory input for exclusive Vcb determinations}

\ShortTitle{The role of theory input for exclusive Vcb determinations}

\author{\speaker{Stefan Schacht}  \\
        Dipartimento di Fisica, Universit\`a di Torino \& INFN, Sezione di Torino, I-10125 Torino, Italy 
	E-mail: \email{schacht@to.infn.it}}

\abstract{
Available form factor parametrizations for $B\rightarrow D^*l\nu$ imply different theoretical assumptions and different treatments of theoretical uncertainties. They give results for $\vert V_{cb}\vert$ whose central values are apart by up to $8\%$. The way the Caprini Lellouch Neubert (CLN) parametrization has been used in experimental analyses sets theoretical uncertainties of the Heavy Quark Effective Theory~(HQET) results on slope and curvature of the form factor ratios $R_1$ and $R_2$ to zero. Furthermore, the relation of curvature and slope of the axial form factor $A_1$ is fixed to the HQET central value. In view of the current experimental precision these uncertainties cannot be neglected any more. 
Using the Boyd Grinstein Lebed~(BGL) parametrization and taking into account theoretical uncertainties in a conservative way, we extract $\vert V_{cb}\vert$ from recent preliminary Belle data and the world average of the total branching ratio. We include an $\mathcal{O}(10\%-20\%)$ theoretical uncertainty of HQET input due to unknown corrections beyond NLO which were neglected in all previous analyses. This is important for reliable extractions of $\vert V_{cb}\vert$ as well as precision tests of the Standard Model with robust predictions of the lepton flavor nonuniversality observable $R(D^*)$ and the $\tau$ polarization asymmetry $P_{\tau}$. Including input from Light Cone Sum Rules (LCSRs) we find $\vert V_{cb}\vert = 40.6\left(^{+1.2}_{-1.3}\right)\cdot 10^{-3}$, $R(D^*) = 0.260(8)$ and $P_{\tau}=-0.47(4)$. Without LCSRs we find $\vert V_{cb}\vert = 41.5(1.3)\cdot 10^{-3}$ and the same results for $R(D^*)$ and $P_{\tau}$. The $R(D^*)$ anomaly is persistent, but its statistical significance is slightly reduced to 2.6$\sigma$.
}

\FullConference{The European Physical Society Conference on High Energy Physics\\
                 5-12 July\\
                 Venice, Italy}

\begin{document}

\section{Introduction}

$V_{cb}$ is an element of the Cabibbo-Kobayashi-Maskawa (CKM) quark mixing matrix and as such a fundamental parameter of the Standard Model (SM). 
It plays an important role for the search for New Physics (NP) in global fits that overconstrain the Unitarity Triangle \cite{Bona:2006ah, Hocker:2001xe}. 
The ratio $\vert V_{ub}/V_{cb}\vert$ directly constrains one side of the CKM triangle. 
Different methods for the extraction of $V_{cb}$ show long-standing discrepancies. The Heavy Flavor Averaging Group (HFLAV) summarizes the current situation as \cite{Amhis:2016xyh}
\begin{align}
\vert V_{cb}\vert &= ( 42.19 \pm 0.78 )\cdot 10^{-3}                                        & \text{from} \quad B &\rightarrow X_c l\nu_l\,, \\
\vert V_{cb}\vert &= ( 39.05 \pm 0.47_{\mathrm{exp}} \pm 0.58_{\mathrm{th}}) \cdot 10^{-3}  & \text{from} \quad B &\rightarrow D^*l\nu_l\,,  \\
\vert V_{cb}\vert &= ( 39.18 \pm 0.94_{\mathrm{exp}} \pm 0.36_{\mathrm{th}} ) \cdot 10^{-3} & \text{from} \quad B &\rightarrow Dl\nu_l\,,    
\end{align}
where $l=e,\mu$. For a discussion of $B\rightarrow Dl\nu$ see also Ref.~\cite{Bigi:2016mdz}.
A key issue in the extraction of $\vert V_{cb}\vert$ is that we have only a limited knowledge of the hadronic form factors. 
Recently, Belle published new preliminary $B\rightarrow D^*l\nu_l$ data which is independent of a certain form factor 
parametrization~\cite{Abdesselam:2017kjf}. This triggered 
a lot of new theoretical studies~\cite{Bernlochner:2017jka, Bigi:2017njr, Grinstein:2017nlq, Bigi:2017jbd, Jaiswal:2017rve,Bernlochner:2017xyx}.
Here, we present the results of our recent work Refs.~\cite{Bigi:2017njr, Bigi:2017jbd}. 
We discuss the available theoretical form factor constraints and parametrizations in Sec.~\ref{sec:theory}. 
Results for $\vert V_{cb}\vert$ are shown in Sec.~\ref{sec:Vcb}. 
After that, in Sec.~\ref{sec:RDstar} we give predictions for the 
observables $R(D^*)$ and $P_{\tau}$, which can be used for precision tests of the SM. To conclude, we briefly summarize our results. 

\section{Theory Constraints on Form Factors \label{sec:theory}}

In the limit of massless leptons the decay $B\rightarrow D^*l\nu$ depends on two axial and one vector form factor, which are denoted as 
$A_{1,5}$ and $V_4$, respectively. In order to treat $B\rightarrow D^*\tau \nu_{\tau}$ decays, one needs the additional pseudoscalar form factor $P_1$. 
We adopt here the notation of Ref~\cite{Caprini:1997mu}, see Ref.~\cite{Bigi:2017jbd} for a translation table to the notation of Ref.~\cite{Boyd:1997kz}. 
The form factors can be written as functions of the dimensionless kinematical quantity 
$w   = (m_B^2+m_{D^*}^2-q^2)/(2 m_B m_{D^*})$, where $q^2\equiv(p_B-p_{D^*})^2$.
Dispersion relations allow to relate the semileptonic region $m_l^2 \leq q^2 \leq (m_B-m_{D^*})^2$ to the 
pair-production region beyond threshold $q^2 \geq (m_B+m_{D^*})^2$.
Using perturbative QCD \cite{Grigo:2012ji}, one can constrain the form factors in the pair-production region.
Then, one can translate this constraint back to the semileptonic region using analyticity.
This motivates the model independent Boyd Grinstein Lebed (BGL) parametrization~\cite{Boyd:1997kz, Boyd:1994tt, Boyd:1995cf}, which 
performs the form factor expansion
\begin{align}
F(z) &= \frac{1}{P_F(z) \phi_F(z)} \sum_{n=0}^{\infty} a^F_n z^n\,, & 
z \equiv \frac{\sqrt{1+w}-\sqrt{2}}{\sqrt{1+w} + \sqrt{2}}\,, \label{eq:BGL} 
\end{align} 
with the outer function $\phi(z)$, the Blaschke factor $P(z)$ and the expansion coefficients $a_n^F$, for details see Ref.~\cite{Boyd:1997kz}.
The BGL parameters $a_n^F$ are bounded by the unitarity conditions \cite{Boyd:1997kz}
\begin{align}
\sum_{n=0}^{\infty} \left(a_n^{V_4}\right)^2 &\leq 1\,, &
\sum_{n=0}^{\infty} \left( \left(a_n^{A_1}\right)^2 + \left(a_n^{A_5}\right)^2 \right) &\leq 1\,. \label{eq:weakunitarity} 
\end{align}
The unitarity bounds Eq.~(\ref{eq:weakunitarity}) are the (weak) special case of the general (strong) unitarity conditions  
which include also the contributions from the BGL parameters of all other $b\rightarrow c$ channels, 
such as $B\rightarrow D$, $B^*\rightarrow D$, and $B^*\rightarrow D^*$~\cite{Boyd:1997kz}.  
For $B\rightarrow D^*l\nu$ the expansion parameter $z$ lies in the range $0<z<0.056$.
This and the unitarity bounds Eqs.~(\ref{eq:weakunitarity}) imply that the expansions Eq.~(\ref{eq:BGL}) converge very fast.
We have already $z^3\sim 10^{-4}$, so that in practice taking into account exponents up to the power of two is already enough. 

Additional information on the form factors is provided by Lattice QCD (LQCD), Heavy Quark Effective Theory (HQET) and 
Light Cone Sum Rules (LCSRs).
LQCD provides the normalization for the $\vert V_{cb}\vert$ extraction with the form factor value $A_1(w=1) =0.902(12)$
(our average from Refs.~\cite{Bailey:2014tva, Harrison:2016gup}).
LCSRs give values at the other end of the kinematic spectrum:  
$A_1(w_{max}) = 0.65(18)$, 
$R_1(w_{max}) = 1.32(4) $, and  
$R_2(w_{max}) = 0.91(17)$ \cite{Faller:2008tr}.
We will show fit results with and without including LCSR input.
HQET and QCD sum rules \cite{Bernlochner:2017jka, Caprini:1997mu, Luke:1990eg, Neubert:1991xw, Neubert:1993mb, Ligeti:1993hw, Neubert:1992pn, Neubert:1992wq} 
give strong constraints for all the $B^{(*)}\rightarrow D^{(*)}$ form factors. In the heavy quark limit $m_{c,b}\gg \Lambda_{\mathrm{QCD}}$ all 
of them can be written using a single Isgur-Wise function. NLO corrections at $\mathcal{O}(\Lambda_{\mathrm{QCD}}/{m_{c,b}}, \alpha_s)$  
are known and can be written in terms of three subleading Isgur-Wise functions.
Following the calculation of Ref.~\cite{Bernlochner:2017jka} we updated all $B^{(*)}\rightarrow D^{(*)}$ form factor ratios, 
see Table~II in Ref.~\cite{Bigi:2017jbd}, which updates Table~A.1 in Ref.~\cite{Caprini:1997mu}. 
The parametric error of the NLO contributions can be taken into account by varying the corresponding subleading parameters as given in Ref.~\cite{Bernlochner:2017jka}.
We also take into account the theoretical uncertainty due to the unknown corrections beyond NLO, which are 
parametrically $\mathcal{O}(\alpha_s^2, \Lambda_{\mathrm{QCD}}^2/m_{c,b}^2, \alpha_s \Lambda_{\mathrm{QCD}}/m_{c,b})$. 
A reliable estimate of their size is complicated by the fact that at zero recoil several form factors are protected from NLO power corrections 
through Luke's theorem~\cite{Luke:1990eg}, which does not apply to the N$^2$LO corrections. 
The form factors which are not protected by Luke's theorem do have NLO corrections up to 60\%. Actually, we have~\cite{Bigi:2017jbd}
\begin{align}
\frac{V_6(w)}{V_1(w)} &= 1.0\,,  & (\text{LO}) \\ 
\frac{V_6(w)}{V_1(w)} &= 1.58 ( 1 -0.18 (w-1) + \dots )\,. & (\text{NLO}) 
\end{align}
For an estimate it is also instructive to compare LQCD and HQET results: 
{\allowdisplaybreaks\begin{align}
 \frac{S_1(w)}{V_1(w)}\Big|_{\rm LQCD} &=  0.975(6) +0.055(18) w_1+\dots, &
\left. \frac{S_1(w)}{V_1(w)} \right|_{\text{HQET}} &= 1.021(30) - 0.044(64) w_1 +\dots \nonumber\\
 \frac{A_1(1)}{V_1(1)}\Big|_{\rm LQCD} &=  0.857(15), & 
\left. \frac{A_1(1)}{V_1(1)} \right|_{\text{HQET}} &= 0.966(28) \nonumber\\
 \frac{S_1(1)}{A_1(1)}\Big|_{\rm LQCD} &=  1.137(21), &
\frac{S_1(1)}{A_1(1)}\Big|_{\rm HQET}  &=  1.055(2), \nonumber
 \end{align}}
where $w_1=w-1$. Between the LQCD and HQET results there are deviations of $5\%-13\%$ which must come from higher order corrections beyond NLO. 

Taking everything into account, N$^2$LO corrections as large as $\mathcal{O}(10\%-20\%)$ cannot be excluded for robust tests of the SM and 
reliable extractions of~$V_{cb}$.

Using the results from HQET, it is possible to relate the BGL parameters of the other $B^{(*)}\rightarrow D^{(*)}$ modes to the ones of $B\rightarrow D^*$. 
In this way we derive the strong version of the unitarity constraints Eq.~(\ref{eq:weakunitarity}), one for each Lorentz structure.
In order to be conservative, in the derivation of these strong unitarity constraints we allow for deviations from the central value of the HQET result 
by $\pm 25\%$ ($\pm 30\%$) at zero (maximal) recoil, which includes both NLO and N$^2$LO corrections. 
We use these constraints as a side condition in the fit. 
A different method to utilize the strong unitarity relations is to eliminate directly some of the form factor parameters and to obtain in this way a 
simplified form factor parametrization. Of course, the theoretical uncertainty of this operation has to be taken into account.
This is the strategy of the Caprini Lellouch Neubert (CLN) parametrization~\cite{Caprini:1997mu}.
A form of this parametrization which is traditionally used in experimental analyses is (see \emph{e.g.}~Ref.~\cite{Abdesselam:2017kjf})
\begin{align}
h_{A_1}(w) &= h_{A_1}(1) \left( 1 - 8 \rho^2 z + (53 \rho^2 - 15 ) z^2 - (231 \rho^2 - 91 ) z^3\right)\,, \label{eq:CLN-1}\\
R_1(w)     &= R_1(1) - 0.12 (w - 1) + 0.05 (w - 1)^2 \,, 						  \label{eq:CLN-2}\\		
R_2(w)     &= R_2(1) + 0.11 (w - 1) - 0.06 (w - 1)^2 \,,					   	  \label{eq:CLN-3}
\end{align}
where $R_1\equiv V_4/A_1$ and $R_2$ is related to the form factor ratio $A_5/A_1$. 
Note that in the experimental analyses the slope and curvature of $R_1$ and $R_2$ as well as the relation of slope and curvature of $h_{A_1}$ are kept fixed,~\emph{i.e.},~parts of the theoretical uncertainties of HQET which were noted in Ref.~\cite{Caprini:1997mu} are neglected \cite{Bernlochner:2017jka, Bigi:2017njr, Grinstein:2017nlq}.

\section{Extraction of $V_{cb}$ \label{sec:Vcb}}

\begin{table}[t]
\begin{center} 
\begin{tabular}{ccccccc}
\hline
\hline
Fit & BGL weak & BGL weak & BGL strong & BGL strong & CLN & CLN \\ 
LCSR                 & $\times$                        & \checkmark &  $\times$  & \checkmark  & $\times$   & $\checkmark$   \\
$\chi^2/\mathrm{dof}$ 	 & 28.2/33			&	32.0/36			& 29.6/33		  & 33.1/36	       & 35.4/37   &  35.9/40  \\ 
$\vert V_{cb}\vert$      & $0.0424\left(18\right)$      & $0.0413\left(14\right)$       & $0.0415\left(13\right)$  & $0.0406\left(^{+12}_{-13}\right)$ &$0.0393(12)$      & $0.0392(12)$\\ 
\hline\hline
\end{tabular}
\caption{Extractions of $\vert V_{cb}\vert$ using BGL and CLN parametrizations with and without LCSR input.
For BGL we also show fit results with and without strong unitarity constraints.
Table adapted and extended from Refs.~\cite{Bigi:2017njr, Bigi:2017jbd}. \label{tab:fit}}
\end{center} 
\end{table}

Summarizing the discussions of Sec.~\ref{sec:theory}, we distinguish three ways to treat the $B\rightarrow D^*$ form factors:
(1) BGL using only weak unitarity, 
(2) BGL using strong unitarity as an additional constraint in the fit, and 
(3) CLN, which uses strong unitarity to obtain a simplified parametrization. 
We fit the BGL and CLN parameters to the global HFLAV average of the total branching 
ratio $\mathcal{B}(\bar{B}^0\rightarrow D^{*+}l^- \bar{\nu}_l) = 0.0488 \pm 0.0010$ \cite{Amhis:2016xyh} 
and the recent preliminary Belle data~\cite{Abdesselam:2017kjf}. 
We use the CLN parametrization in the traditional form of Eqs.~(\ref{eq:CLN-1})--(\ref{eq:CLN-3}).
The results are shown in Table~\ref{tab:fit}. 
Our fit scenario (1) agrees very well with Ref.~\cite{Grinstein:2017nlq}.
Note that while all of the fits are compatible with each other, the central values for $\vert V_{cb}\vert$ extracted using the BGL and 
the CLN parametrization differ by up to 7.9\%, and 5.4\% including the LCSR input. 
Including the strong unitarity bounds, the deviation is reduced to 5.6\% and 3.6\% (with LCSR).
The main reason for the deviation is that the CLN parametrization Eqs.~(\ref{eq:CLN-1})--(\ref{eq:CLN-3}) does not take into account any 
theoretical uncertainties for the slope of $R_{1}$ and $R_{2}$, but fixes them to the central value of the HQET result. 
Consequently, the BGL parametrization is more flexible, especially near $w=1$. 
This is the most important kinematical region of the data, because here we have LQCD 
information on the normalization of the form factor $A_1$, which is essential for the extraction of $V_{cb}$.
Relaxing the bound on $R_1$ and $R_2$ in the CLN parametrization leads to a result which is quite comparable to the BGL 
fits, see Refs.~\cite{Bigi:2017njr, Bernlochner:2017xyx}.

\section{SM Predictions for $R(D^*)$ and $P_{\tau}$ \label{sec:RDstar}}

\begin{table}[t]
\begin{center} 
\begin{tabular}{ccccccc}
\hline\hline
Ref.	   & Our result \cite{Bigi:2017jbd} & \cite{Fajfer:2012vx}  & \cite{Celis:2012dk}		  &\cite{Bernlochner:2017jka}	& \cite{Jaiswal:2017rve}  & \cite{Tanaka:2012nw,Abdesselam:2016xqt} \\
$R(D^*)$   & 0.260(8)  & 0.252(3)  & $0.252(2)(3)$			      & 0.257(3)  & 0.257(5)      	&  0.252(4) \\
$P_{\tau}$ & $-0.47(4)$  &	   & $-0.502(^{+5}_{-6})(17)$ &		  & 			&  $-0.497(13)$ \\
Deviation  & $2.6\sigma$ & $3.5\sigma$ & $3.4\sigma$ & $3.1\sigma$ & $3.0\sigma$ & $3.4\sigma$  \\\hline\hline
\end{tabular}
\caption{Our results for $R(D^*)$ and $P_{\tau}$ compared with other theoretical results available in the literature. 
The experimental measurements are $R(D^*)^{\mathrm{exp}} = 0.304(13)(7)$ \cite{Amhis:2016xyh} and $P_{\tau}^{\mathrm{exp}} = -0.38(51)(^{+21}_{-16})$ \cite{Hirose:2016wfn,Hirose:2017dxl}. The last line shows the respective deviation from the measurement $R(D^*)^{\mathrm{exp}}$. 
\label{tab:Rdstar}}
\end{center} 
\end{table}

In order to predict $R(D^*) \equiv \mathcal{B}(B\rightarrow D^*\tau\nu) / \mathcal{B}(B\rightarrow D^*l\nu)$ and the $\tau$ polarization asymmetry 
$P_{\tau} \equiv (\Gamma^+-\Gamma^-)/(\Gamma^++\Gamma^-)$, where $\Gamma^{\pm}$ are the integrated polarized decay rates, we need theory input for the pseudoscalar form 
factor $P_1$. This form factor is not constrained by $B\rightarrow D^*l\nu$ data. 
Employing its BGL parametrization we use the three constraints (1)~strong unitarity, (2)~the kinematical endpoint relation $P_1(w_{\mathrm{max}}) = A_5(w_{\mathrm{max}})$ with our fit result for $A_5(w_{\mathrm{max}})$, and (3)~the HQET result $P_1(1) = 1.21\pm 0.06 \pm 0.18$, where the first error is the parametric NLO error and the second error is our estimate of the N$^2$LO uncertainty as $15\%$ of the central value, see our discussion in Sec.~\ref{sec:theory}.
These conditions determine the three $P_1$ BGL parameters. Our results are shown in Table~\ref{tab:Rdstar}. 
They are consistent with the results in the literature, however the value of $R(D^*)$ and the uncertainties that we obtain are larger. 
We find a $2.6\sigma$ deviation from the experimental measurement of~$R(D^*)^{\mathrm{exp}}$.

\section{Conclusions}

We use recent preliminary Belle data on $B\rightarrow D^*l\nu$ which is independent of a particular form factor parametrization in order 
to reappraise the methodology of exclusive $V_{cb}$ extractions and predictions of $R(D^*)$ and $P_{\tau}$. 
It turns out that the BGL parametrization and the form of the CLN parametrization as used in experimental analyses give different results for $V_{cb}$.
This stays true when one includes strong unitarity constraints as external constraints on the BGL parameters. 
The reason is that the used form of the CLN parametrization neglects important theoretical 
uncertainties which leads to less flexibility of the parametrization near $w=1$, where LQCD gives information on the form factor normalization, 
which is essential for the determination of $\vert V_{cb}\vert$. 
We advocate therefore to use the BGL parametrization with the strong unitarity constraints as side conditions, taking into account 
the input from HQET in a conservative way. 
We estimate the theoretical uncertainty of the HQET input due to unknown higher order corrections to be $\mathcal{O}(10\%-20\%)$ on top 
of the parametric NLO error. Our results for $\vert V_{cb}\vert$, $R(D^*)$ and $P_{\tau}$ are given in Tables~\ref{tab:fit} and \ref{tab:Rdstar}.
The $R(D^*)$ anomaly is persistent, but slightly reduced to $2.6\sigma$. 
 
As we rely except for the world average of the total branching ratio on recent, preliminary data, we have to be patient: the $V_{cb}$ puzzle is not yet solved. 
A reanalysis of previous BaBar and Belle data is necessary, taking properly into account theoretical uncertainties.
Together with future lattice data which determines the slope of the form factors this will conclusively settle the case.

\section*{Acknowledgments}
StS is happy to thank his collaborators Dante Bigi and Paolo Gambino. 

\bibliographystyle{JHEP}
\bibliography{draft.bib}

\providecommand{\href}[2]{#2}\begingroup\raggedright\begin{thebibliography}{10}

\bibitem{Bona:2006ah}
{\scshape UTfit} collaboration, M.~Bona et~al., \emph{{The Unitarity Triangle
  Fit in the Standard Model and Hadronic Parameters from Lattice QCD: A
  Reappraisal after the Measurements of $\Delta m_s$ and BR($B \rightarrow \tau
  \nu_{\tau}$)}},
  \href{https://doi.org/10.1088/1126-6708/2006/10/081}{\emph{JHEP} {\bfseries
  10} (2006) 081}, [\href{https://arxiv.org/abs/hep-ph/0606167}{{\ttfamily
  hep-ph/0606167}}].

\bibitem{Hocker:2001xe}
A.~Hocker, H.~Lacker, S.~Laplace and F.~Le~Diberder, \emph{{A New approach to a
  global fit of the CKM matrix}},
  \href{https://doi.org/10.1007/s100520100729}{\emph{Eur. Phys. J.} {\bfseries
  C21} (2001) 225--259},
  [\href{https://arxiv.org/abs/hep-ph/0104062}{{\ttfamily hep-ph/0104062}}].

\bibitem{Amhis:2016xyh}
Y.~Amhis et~al., \emph{{Averages of $b$-hadron, $c$-hadron, and $\tau$-lepton
  properties as of summer 2016, and online update at
  http://www.slac.stanford.edu/xorg/hflav}},
  \href{https://arxiv.org/abs/1612.07233}{{\ttfamily 1612.07233}}.

\bibitem{Bigi:2016mdz}
D.~Bigi and P.~Gambino, \emph{{Revisiting $B\to D \ell \nu$}},
  \href{https://doi.org/10.1103/PhysRevD.94.094008}{\emph{Phys. Rev.}
  {\bfseries D94} (2016) 094008},
  [\href{https://arxiv.org/abs/1606.08030}{{\ttfamily 1606.08030}}].

\bibitem{Abdesselam:2017kjf}
{\scshape Belle} collaboration, A.~Abdesselam et~al., \emph{{Precise
  determination of the CKM matrix element $\left| V_{cb}\right|$ with $\bar B^0
  \to D^{*\,+} \, \ell^- \, \bar \nu_\ell$ decays with hadronic tagging at
  Belle}},  \href{https://arxiv.org/abs/1702.01521}{{\ttfamily 1702.01521}}.

\bibitem{Bernlochner:2017jka}
F.~U. Bernlochner, Z.~Ligeti, M.~Papucci and D.~J. Robinson, \emph{{Combined
  analysis of semileptonic $B$ decays to $D$ and $D^*$: $R(D^{(*)})$,
  $|V_{cb}|$, and new physics}},
  \href{https://doi.org/10.1103/PhysRevD.95.115008}{\emph{Phys. Rev.}
  {\bfseries D95} (2017) 115008},
  [\href{https://arxiv.org/abs/1703.05330}{{\ttfamily 1703.05330}}].

\bibitem{Bigi:2017njr}
D.~Bigi, P.~Gambino and S.~Schacht, \emph{{A fresh look at the determination of
  $|V_{cb}|$ from $B\to D^{*} \ell \nu$}},
  \href{https://doi.org/10.1016/j.physletb.2017.04.022}{\emph{Phys. Lett.}
  {\bfseries B769} (2017) 441--445},
  [\href{https://arxiv.org/abs/1703.06124}{{\ttfamily 1703.06124}}].

\bibitem{Grinstein:2017nlq}
B.~Grinstein and A.~Kobach, \emph{{Model-Independent Extraction of $|V_{cb}|$
  from $\bar{B}\rightarrow D^* \ell \overline{\nu}$}},
  \href{https://doi.org/10.1016/j.physletb.2017.05.078}{\emph{Phys. Lett.}
  {\bfseries B771} (2017) 359--364},
  [\href{https://arxiv.org/abs/1703.08170}{{\ttfamily 1703.08170}}].

\bibitem{Bigi:2017jbd}
D.~Bigi, P.~Gambino and S.~Schacht, \emph{{$R(D^*)$, $|V_{cb}|$, and the Heavy
  Quark Symmetry relations between form factors}},
  \href{https://arxiv.org/abs/1707.09509}{{\ttfamily 1707.09509}}.

\bibitem{Jaiswal:2017rve}
S.~Jaiswal, S.~Nandi and S.~K. Patra, \emph{{Extraction of $|V_{cb}|$ from
  $B\to D^{(*)}\ell\nu_\ell$ and the Standard Model predictions of
  $R(D^{(*)})$}},  \href{https://arxiv.org/abs/1707.09977}{{\ttfamily
  1707.09977}}.

\bibitem{Bernlochner:2017xyx}
F.~U. Bernlochner, Z.~Ligeti, M.~Papucci and D.~J. Robinson, \emph{{Tensions
  and correlations in $|V_{cb}|$ determinations}},
  \href{https://arxiv.org/abs/1708.07134}{{\ttfamily 1708.07134}}.

\bibitem{Caprini:1997mu}
I.~Caprini, L.~Lellouch and M.~Neubert, \emph{{Dispersive bounds on the shape
  of $\bar{B}\rightarrow D^{(*)} l\bar{\nu}$ form-factors}},
  \href{https://doi.org/10.1016/S0550-3213(98)00350-2}{\emph{Nucl. Phys.}
  {\bfseries B530} (1998) 153--181},
  [\href{https://arxiv.org/abs/hep-ph/9712417}{{\ttfamily hep-ph/9712417}}].

\bibitem{Boyd:1997kz}
C.~G. Boyd, B.~Grinstein and R.~F. Lebed, \emph{{Precision corrections to
  dispersive bounds on form-factors}},
  \href{https://doi.org/10.1103/PhysRevD.56.6895}{\emph{Phys. Rev.} {\bfseries
  D56} (1997) 6895--6911},
  [\href{https://arxiv.org/abs/hep-ph/9705252}{{\ttfamily hep-ph/9705252}}].

\bibitem{Grigo:2012ji}
J.~Grigo, J.~Hoff, P.~Marquard and M.~Steinhauser, \emph{{Moments of heavy
  quark correlators with two masses: exact mass dependence to three loops}},
  \href{https://doi.org/10.1016/j.nuclphysb.2012.07.007}{\emph{Nucl. Phys.}
  {\bfseries B864} (2012) 580--596},
  [\href{https://arxiv.org/abs/1206.3418}{{\ttfamily 1206.3418}}].

\bibitem{Boyd:1994tt}
C.~G. Boyd, B.~Grinstein and R.~F. Lebed, \emph{{Constraints on form-factors
  for exclusive semileptonic heavy to light meson decays}},
  \href{https://doi.org/10.1103/PhysRevLett.74.4603}{\emph{Phys. Rev. Lett.}
  {\bfseries 74} (1995) 4603--4606},
  [\href{https://arxiv.org/abs/hep-ph/9412324}{{\ttfamily hep-ph/9412324}}].

\bibitem{Boyd:1995cf}
C.~G. Boyd, B.~Grinstein and R.~F. Lebed, \emph{{Model independent extraction
  of $|V_{cb}|$ using dispersion relations}},
  \href{https://doi.org/10.1016/0370-2693(95)00480-9}{\emph{Phys. Lett.}
  {\bfseries B353} (1995) 306--312},
  [\href{https://arxiv.org/abs/hep-ph/9504235}{{\ttfamily hep-ph/9504235}}].

\bibitem{Bailey:2014tva}
{\scshape Fermilab Lattice, MILC} collaboration, J.~A. Bailey et~al.,
  \emph{{Update of $|V_{cb}|$ from the $\bar{B}\to D^*\ell\bar{\nu}$ form
  factor at zero recoil with three-flavor lattice QCD}},
  \href{https://doi.org/10.1103/PhysRevD.89.114504}{\emph{Phys. Rev.}
  {\bfseries D89} (2014) 114504},
  [\href{https://arxiv.org/abs/1403.0635}{{\ttfamily 1403.0635}}].

\bibitem{Harrison:2016gup}
J.~Harrison, C.~Davies and M.~Wingate, \emph{{$|V_{cb}|$ from the $\bar{B}^0
  \to D^{*+} \ell^- \bar{\nu}$ zero-recoil form factor using $2+1+1$ flavour
  HISQ and NRQCD}}, {\emph{PoS} {\bfseries LATTICE2016} (2017) 287},
  [\href{https://arxiv.org/abs/1612.06716}{{\ttfamily 1612.06716}}].

\bibitem{Faller:2008tr}
S.~Faller, A.~Khodjamirian, C.~Klein and T.~Mannel, \emph{{$B\rightarrow
  D^{(*)}$ Form Factors from QCD Light-Cone Sum Rules}},
  \href{https://doi.org/10.1140/epjc/s10052-009-0968-4}{\emph{Eur. Phys. J.}
  {\bfseries C60} (2009) 603--615},
  [\href{https://arxiv.org/abs/0809.0222}{{\ttfamily 0809.0222}}].

\bibitem{Luke:1990eg}
M.~E. Luke, \emph{{Effects of subleading operators in the heavy quark effective
  theory}}, \href{https://doi.org/10.1016/0370-2693(90)90568-Q}{\emph{Phys.
  Lett.} {\bfseries B252} (1990) 447--455}.

\bibitem{Neubert:1991xw}
M.~Neubert and V.~Rieckert, \emph{{New approach to the universal form-factors
  in decays of heavy mesons}},
  \href{https://doi.org/10.1016/0550-3213(92)90080-U}{\emph{Nucl. Phys.}
  {\bfseries B382} (1992) 97--119}.

\bibitem{Neubert:1993mb}
M.~Neubert, \emph{{Heavy quark symmetry}},
  \href{https://doi.org/10.1016/0370-1573(94)90091-4}{\emph{Phys. Rept.}
  {\bfseries 245} (1994) 259--396},
  [\href{https://arxiv.org/abs/hep-ph/9306320}{{\ttfamily hep-ph/9306320}}].

\bibitem{Ligeti:1993hw}
Z.~Ligeti, Y.~Nir and M.~Neubert, \emph{{The Subleading Isgur-Wise form-factor
  $\xi_3(v\cdot v')$ and its implications for the decays $\bar{B}\rightarrow
  D^* l \bar{\nu}$}},
  \href{https://doi.org/10.1103/PhysRevD.49.1302}{\emph{Phys. Rev.} {\bfseries
  D49} (1994) 1302--1309},
  [\href{https://arxiv.org/abs/hep-ph/9305304}{{\ttfamily hep-ph/9305304}}].

\bibitem{Neubert:1992pn}
M.~Neubert, Z.~Ligeti and Y.~Nir, \emph{{The Subleading Isgur-Wise form-factor
  $\chi_3(v\cdot v')$ to order $\alpha_s$ in QCD sum rules}},
  \href{https://doi.org/10.1103/PhysRevD.47.5060}{\emph{Phys. Rev.} {\bfseries
  D47} (1993) 5060--5066},
  [\href{https://arxiv.org/abs/hep-ph/9212266}{{\ttfamily hep-ph/9212266}}].

\bibitem{Neubert:1992wq}
M.~Neubert, Z.~Ligeti and Y.~Nir, \emph{{QCD sum rule analysis of the
  subleading Isgur-Wise form-factor $\chi_2(v\cdot v')$}},
  \href{https://doi.org/10.1016/0370-2693(93)90728-Z}{\emph{Phys. Lett.}
  {\bfseries B301} (1993) 101--107},
  [\href{https://arxiv.org/abs/hep-ph/9209271}{{\ttfamily hep-ph/9209271}}].

\bibitem{Fajfer:2012vx}
S.~Fajfer, J.~F. Kamenik and I.~Nisandzic, \emph{{On the $B \to D^* \tau \bar
  \nu_{\tau}$ Sensitivity to New Physics}},
  \href{https://doi.org/10.1103/PhysRevD.85.094025}{\emph{Phys. Rev.}
  {\bfseries D85} (2012) 094025},
  [\href{https://arxiv.org/abs/1203.2654}{{\ttfamily 1203.2654}}].

\bibitem{Celis:2012dk}
A.~Celis, M.~Jung, X.-Q. Li and A.~Pich, \emph{{Sensitivity to charged scalars
  in $\boldsymbol{B\to D^{(*)}\tau\nu_\tau}$ and
  $\boldsymbol{B\to\tau\nu_\tau}$ decays}},
  \href{https://doi.org/10.1007/JHEP01(2013)054}{\emph{JHEP} {\bfseries 01}
  (2013) 054}, [\href{https://arxiv.org/abs/1210.8443}{{\ttfamily 1210.8443}}].

\bibitem{Tanaka:2012nw}
M.~Tanaka and R.~Watanabe, \emph{{New physics in the weak interaction of $\bar
  B\to D^{(*)}\tau\bar\nu$}},
  \href{https://doi.org/10.1103/PhysRevD.87.034028}{\emph{Phys. Rev.}
  {\bfseries D87} (2013) 034028},
  [\href{https://arxiv.org/abs/1212.1878}{{\ttfamily 1212.1878}}].

\bibitem{Abdesselam:2016xqt}
A.~Abdesselam et~al., \emph{{Measurement of the $\tau$ lepton polarization in
  the decay ${\bar B} \rightarrow D^* \tau^- {\bar \nu_{\tau}}$}},
  \href{https://arxiv.org/abs/1608.06391}{{\ttfamily 1608.06391}}.

\bibitem{Hirose:2016wfn}
{\scshape Belle} collaboration, S.~Hirose et~al., \emph{{Measurement of the
  $\tau$ lepton polarization and $R(D^*)$ in the decay $\bar{B} \to D^* \tau^-
  \bar{\nu}_\tau$}},
  \href{https://doi.org/10.1103/PhysRevLett.118.211801}{\emph{Phys. Rev. Lett.}
  {\bfseries 118} (2017) 211801},
  [\href{https://arxiv.org/abs/1612.00529}{{\ttfamily 1612.00529}}].

\bibitem{Hirose:2017dxl}
{\scshape Belle} collaboration, S.~Hirose et~al., \emph{{Measurement of the
  $\tau$ lepton polarization and $R(D^*)$ in the decay $\bar{B} \rightarrow D^*
  \tau^- \bar{\nu}_\tau$ with one-prong hadronic $\tau$ decays at Belle}},
  \href{https://arxiv.org/abs/1709.00129}{{\ttfamily 1709.00129}}.

\end{thebibliography}\endgroup

\end{document}